# Impulse Excitation Diagram as a tool to achieve high energy orbits

Damian Gąska
Faculty of Transport and Aviation Engineering
Silesian University of Technology
Katowice, Poland
damian.gaska@polsl.pl

Jerzy Margielewicz
Faculty of Transport and Aviation Engineering
Silesian University of Technology
Katowice, Poland
jerzy.margielewicz@polsl.pl

Grzegorz Litak
Faculty of Mechanical Engineering
Lublin University of Technology
Lublin, Poland
g.litak@pollub.pl

Piotr Wolszczak
Faculty of Mechanical Engineering
Lublin University of Technology
Lublin, Poland
p.wolszczak@pollub.pl

Sławomir Bucki
Faculty of Transport and Aviation Engineering
Silesian University of Technology
Katowice, Poland
slawomir.bucki@polsl.pl

*Abstract*— **The paper presents the application of a new impulse excitation diagram (IED) to help realize high-energy orbits in nonlinear energy harvesting systems. In the case of non-linearity, we can deal with the occurrence of coexisting solutions and the proposed diagram allows for the use of the impulse excitation method in order to change the solution. For this purpose, the author's IED diagram was presented to determine the characteristics, duration and moment of initiation of the external disturbing impulse in order to jump to another orbit. An application example is the quasizero energy harvester and two different impulse characteristics.**

## I. Introduction

The continuous development of science often leads to solutions that constitute a breakthrough in technology. One of such breakthroughs, practically at the beginning of this century, may be energy harvesting [1]. Research on the recovery of ambient energy in the form of, for example, mechanical vibrations, is currently conducted in many research centers around the world [2,3]. This issue is extended to other sources, mainly such as: air flow (as opposed to energy obtained from wind turbines), temperature changes, [4,5] etc. This arouses increasing interest also among designers, especially in the field of application in IoT devices [6]. The goal, of course, is to eliminate the external power source, which simplifies the operation of the device, eliminates the need to replace the battery or connect the device to the network. The amounts of energy harvested in this way is small, but sufficient to power sensors and transmit information [7].

Devices for harvesting kinetic energy from mechanical vibrations and converting it into electrical energy with use of electrostatic [8], electromagnetic [9,10] or piezoelectric [11,12] methods. The use of piezoelectric transducers has gained quite a lot of popularity due to high energy efficiency. The frequency of vibrations in the human environment is usually quite low and is below 200 Hz, while in the case of infrastructure elements such as bridges, it often does not exceed 5 Hz [13]. The occurring amplitudes of vibrations depend on the object and may vary in a wide range. As a result, various designs of energy harvesting systems are created, which were initially adapted to a specific object. An energy harvester using the piezoelectric phenomenon is most often made of a housing mounted on a vibrating object and a beam excited to vibrations with a piezoelectric attached to it. The deformation of the beam causes the formation of an electric charge in the electric circuit of the piezoelectric [14]. Initially, such systems were designed to be linear, focused on the resonant frequency of the source of the excitation, but the work [15] resulted in a breakthrough by introducing non-linearity. Since then, non-linear broadband energy harvesters have been developed, the operating spectrum of which is much wider than that of devices with linear characteristics [16]. The nonlinearities are most often introduced by a system of magnets resulting in the formation of bistable [16] energy harvesters, tristable [17] and generally multistable [18]. In addition, nonlinearities are introduced by usage of variable beam structures, materials, elastic and dissipative elements [19–21].

Nonlinear energy harvesters, despite their indisputable advantages (broadband work with various input parameters), have a characteristic feature for all nonlinear systems – coexisting solutions may occur in them [22]. For the same excitation conditions, there may be many orbits, some of which will be more energy efficient and others less. Under zero initial conditions, nonlinear energy harvesters usually vibrate at the low-energy orbit, which leads to small energy harvesting efficiency [18]. Of course, the goal is for the energy harvester to work with the greatest possible energy efficiency, therefore research is being carried out on the possibility of controlling the orbit.

Currently, two types of orbit change method are applicable. These are electrical and mechanical methods based on the use of an external impact to achieve a high-energy orbit. Examples of electrical methods are voltage impulse perturbation presented in the paper [23], and load perturbation [24] based on electromagnetic kick. For the second method, an impact-induced spring-based structure was presented by Zhou et al. [25] to achieve high-energy orbits to overcome local potential barriers. Some solutions are based on buckling level modification [26,27], while at work Yan et al. a synchronized switch stiffness control technique was proposed [28]. Similar studies developing the mechanical method were carried out in the works [29,30]. The key question is how much energy will be obtained and how much will be needed for the jump between orbits because the impulse should be powered by energy harvested by the system. The paper presents the IED tool that allows to solve some of the problems related to achieving high energy orbits.

## II. APPLICATION OF IED FOR EXAMPLE ENERGY HARVESTER

### A. Quasi-Zero Energy Harvester

To present the use of IED in practice, at the beginning we present the energy harvesting system designed by us and analyzed for some time – Quasi-Zero Energy Harvester (QZEH) [20,21] in which the potential function was mapped with the quasi-zero stiffness characteristic. The design structure of analyzed device is shown in Fig. 1. In comparison to other design solutions, based on a flexible beam and permanent magnets, the QZEH is characterized by an almost flat energy potential well. The quasizero stiffness subsystem (Fig. 1 – bottom view) consists of two compensation springs and a main spring which are fixed in the housing, and with the other ends in the mass suspended on a flexible cantilever beam. The operation of the system is similar to that of energy harvesters based on permanent magnets, however, in this case, non-linear effects are caused by the use of a system of elastic and dissipative elements. Generally speaking, the deformation of the beam is caused by the external excitation $f$ (for model tests we have assumed harmonic excitation $f = A_1 sin(\omega_1 t) + A_2 sin(\omega_2 t)$) acting on the system through the $IV$ frame, causing the mass $m$ to move, and thus the conversion of kinetic energy into electric energy using piezoelectric $II$.

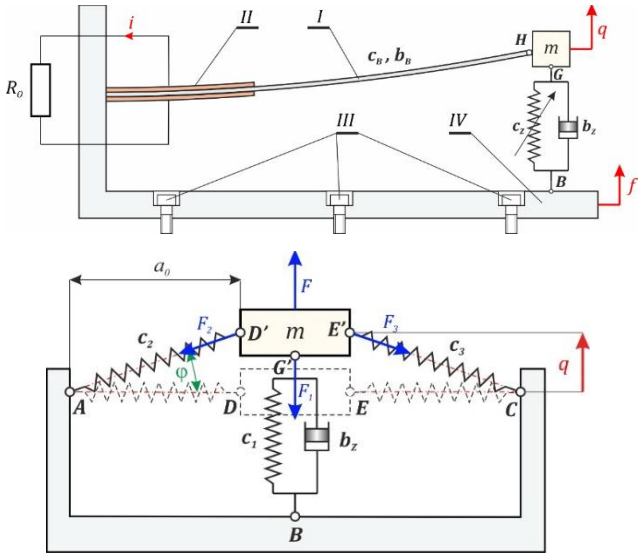

Fig. 1. Phenomenological model of the QZEH energy harvesting system [20]

For a full analysis of the structure and operation of the quasi-zero energy harvester, please read the work [20]. There, the derivation of the mathematical model is presented, the characteristics of the potential function and the influence of individual parameters on its shape are discussed in detail. Here we present only the final form of the dimensionless model. In order to identify all variables, we encourage you to analyze the publications [20,21]. It contains the assumed values, while in this conference paper we decided not to present them due to the limited volume of the paper.

$$\frac{d^2x}{d\tau^2} + \delta\frac{dx}{d\tau} + \left(1 - \frac{1}{\sqrt{1+x^2}}\right)x - \theta U_P = \omega^2 p\left(sin(\omega\tau) + \mu^2 sin(\mu\omega\tau)\right)$$

$$\frac{dU_P}{d\tau} + \sigma U_P + \vartheta\frac{dx}{d\tau} = 0$$

where:

$$\omega_0^2 = \frac{c_Z}{m}, \quad \delta = \frac{b}{m\omega_0}, \quad \theta = \frac{k_P}{ma_0\omega_0^2}, \quad p = \frac{A}{a_0}, \quad \omega = \frac{\omega_1}{\omega_0},$$
$$x = \frac{y}{a_0}, \quad \vartheta = \frac{k_P a_0}{C_P}, \quad \sigma = \frac{1}{\omega_0 C_P R_o}, \quad \mu = \frac{\omega_2}{\omega_1},$$

### B. Coexisting solutions and energy efficiency

In order to present the application of IEDs, it is necessary to conduct computer simulations aimed at identifying individual solutions and their energy efficiency. For this purpose, the results of numerical analyzes are presented in the form of an exemplary basin of attraction (Fig. 2). There we can observe 4 possible solutions that differ in the probability of their occurrence and their periodicity. Examples of energy harvester work parameters were selected in the form of the excitation amplitude $p$=0.4, excitation frequency $\omega$=1.3 and frequency ratio of individual harmonics $\mu$=3/4.

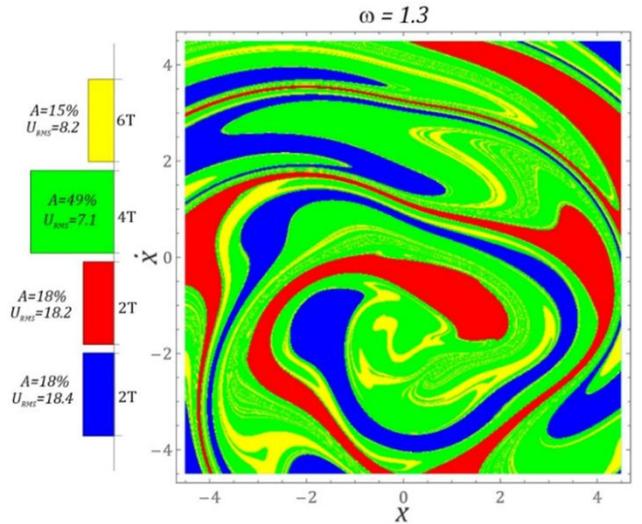

Fig. 2. Basin of attraction of coexisting solutions for excitation amplitude $p$=0.4, excitation frequency $\omega$=1.3 and frequency ratio of individual harmonics $\mu$=3/4 [20]

Next to the basin of attraction there is a bar graph showing the probability of obtaining solutions as a percentage of the basin area, periodicity of each solution and its energy efficiency in the form of RMS voltage $U_{RMS}$. In the analyzed case, there are 4 solutions marked with colors. The most effective are solutions with a periodicity of 2T, of which the solutions marked in blue are characterized by the value of $U_{RMS}$ = 18.4. A slightly lower value occurs for a solution also with a periodicity of 2T and marked in red. The least energy efficiency has a 4T solution marked with green color, whose share in the basin of attraction is the largest. Therefore, there is the greatest probability of obtaining it and also ineffective work of the energy harvester. The amount of energy that can be harvested is more than twice lower than the 2T periodic solutions. In the case of the 6T periodic solution, the situation is similar to the 4T solution. The orbits of individual solutions are shown in Fig. 3. We can observe that the first two 2T periodic solutions are characterized by the largest amplitude.

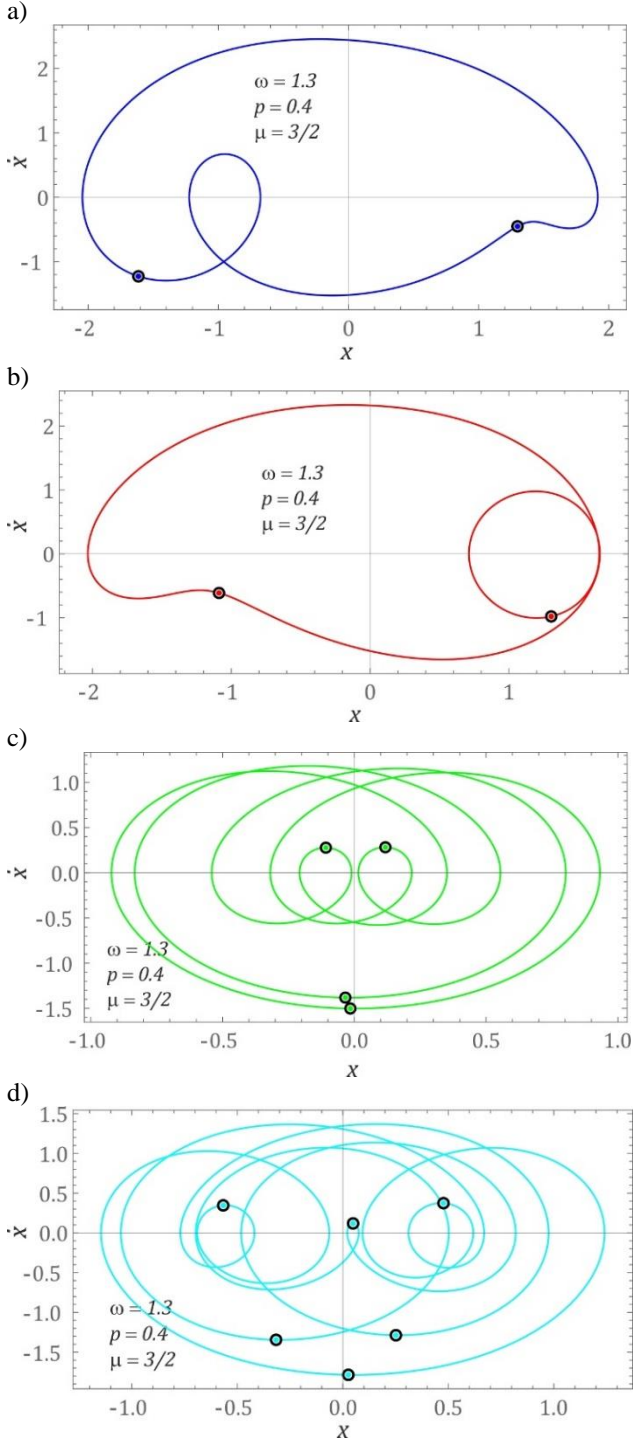

Fig. 3. Orbits of particular solutions for excitation apmlitude $p$=0.4, excitation frequency $\omega$=1.3 and frequency ratio of individual harmonics $\mu$=3/4, a) solution with 2T periodicity and $U_{RMS}$=18.4, b) solution with 2T periodicity and $U_{RMS}$=18.2, c) solution with 4T periodicity and $U_{RMS}$=7.1, d) solution with 6T periodicity and $U_{RMS}$=8.2.

## C. Application of Impulse Excitation Diagram

In the case of multistable energy harvesters, it is necessary to assign such an impulse value to overcome the local barriers of the potential function and achieve an orbit with the largest amplitude possible. To carry out such a task effectively, it is necessary to know the exact duration of the impulse and its amplitude. It is also possible to use a different impulse characteristic, described for example by a half sine wave or a rectangular profile. We can also deal with a sequence of consecutive impulses or their other characteristics, etc. It is also necessary to know the exact time of the impulse initiation depending on the excitation characteristics, so that it is necessary to use as little energy as possible to reach a high energy orbit. For this purpose, the Impulse Excitation Diagram (Fig.4) can be used as a numerical tool supporting and optimizing the procedure.

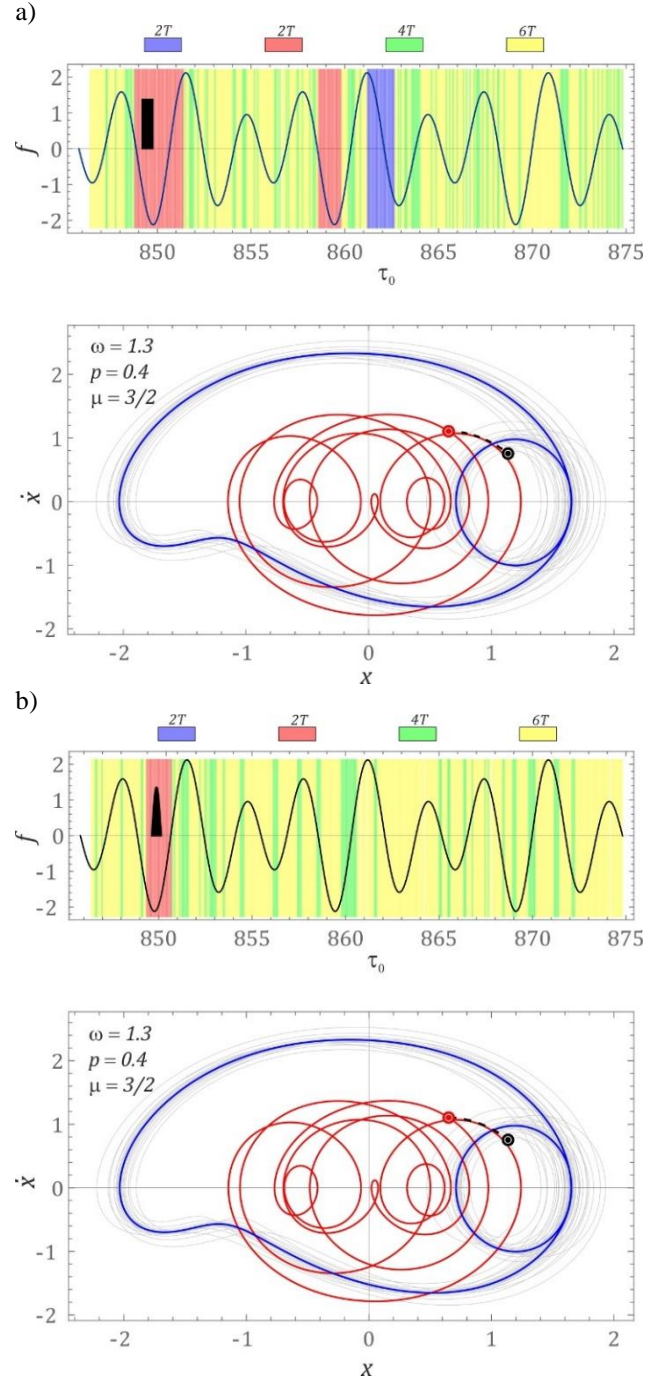

Fig. 4. IED for achiving high energy orbit for excitation apmlitude $p$=0.4, excitation frequency $\omega$=1.3 and frequency ratio of individual harmonics $\mu$=3/4, a) orbit jump with rectangular impuls to solution with 2T periodicity and $U_{RMS}$=18.2, b) orbit jump with half-sinusoidal impuls to solution with 2T periodicity and $U_{RMS}$=18.2.

The diagram is created in the Wolfram Mathematica software using the proprietary script and does not take into account the construction of the impulse excitation system. The construction of such a system and the technical method of impulse activation is not the aim of this paper. The IED consists of two graphs, the first of which shows the time sequence in the form of the excitation amplitude (dark blue) and the shape and duration, amplitude and place of initiation of the disturbing impulse (black) allowing a jump to another orbit. Colored vertical stripes (moments of time) symbolize individual solutions that can be obtained at a given moment in time. The number of stripes (corresponding to the previously indicated solutions on basin of attraction) allows to determine the probability of obtaining a given solution. The diagrams presented in Fig. 4 show the moment of time $\tau_0$ from 845 to 875 – the remaining values repeat cyclically. The initiated impulse (Fig. 4a) with a rectangular characteristic allows to achieve a solution with periodicity of 2T and a predefined value of $U_{RMS}$ = 18.2. The second part of the IED shows a graphical visualization of the jump to the required orbit, where the red point is the impulse initiation and the black point is the target after the impulse duration. In the case shown in Fig. 4b, the difference is the shape of the impulse – here we are dealing with a half-sinusoidal one.

Fig. 4 shows the results of the numerical calculations that were performed for the same amplitude values of the disturbing impulses. Additionally, the same impulse duration was assumed. In the case of rectangular impulse initiation, we are able to achieve all the coexisting solutions. In the case of a half-sinusoidal impulse, obtaining a solution marked in blue is possible when the impulse amplitude increases by 50%, then the first blue lines appear leading to a possibility of reaching high-energy solution.

## III. Conclusions

The use of the Impulse Excitation Diagram allows for a detailed analysis of the impulse duration, moment of its initiation and its amplitude. Such a graph makes it possible to quickly compare the results for various impulse shapes and to optimize its other parameters. This is very important in the case of the implementation of high-energy orbits in energy harvester, which for zero initial conditions do not necessarily work with the highest efficiency. The use of IEDs is possible for different methods of orbit jumps, allowing for graphic visualization of the possibility of achieving the desired solution.

## Acknowledgment

This research was funded by National Science Centre, Poland under the project SHENG-2, No. 2021/40/Q/ST8/00362